# Evaluation of tools for differential gene expression analysis by RNA-seq on a 48 biological replicate experiment

Nicholas J. Schurch[1,2], Pieta Schofield[1,2,3], Marek Gierliński[1,2,3], Christian Cole[1,2], Alexander Sherstnev[1,2], Vijender Singh[3], Nicola Wrobel[6], Karim Gharbi[6], Gordon G. Simpson[4,7,8,9], Tom Owen-Hughes[3,8,10], Mark Blaxter[6,8,11] and Geoffrey J. Barton[2,3,5,8,12]

[1] These authors contributed equally to this work. [2] Division of Computational Biology, [3] Division of Gene Regulation and Expression, [4] Division of Plant Sciences, and [5] Division of Biological Chemistry and Drug Discovery,  College of Life Sciences, University of Dundee, Dow St, Dundee, DD1 5EH, UK,  [6] Edinburgh Genomics, University of Edinburgh, Edinburgh, EH9 3JT, UK, [7] James Hutton Institute, Invergowrie, Dundee, DD2 5DA, Scotland, UK [8] Joint corresponding authors,
[9] g.g.simpson@dundee.ac.uk, [10] t.a.owenhughes@dundee.ac.uk, [11] Mark.Blaxter@ed.ac.uk, [12] g.j.barton@dundee.ac.uk

**ABSTRACT**

An RNA-seq experiment with 48 biological replicates in each of 2 conditions was performed to determine the number of biological replicates ($n_r$) required, and to identify the most effective statistical analysis tools for identifying differential gene expression (DGE). When $n_r = 3$, seven of the nine tools evaluated give true positive rates (TPR) of only 20-40%. For high fold-change genes ($|\log_2 FC| > 2$) the TPR is $> 85\%$. Two tools performed poorly; over- or under-predicting the number of differentially expressed genes. Increasing replication gives a large increase in TPR when considering all DE genes but only a small increase for high fold-change genes. Achieving a $TPR > 85\%$ across all fold-changes requires $n_r > 20$. For future RNA-seq experiments these results suggest $n_r > 6$, rising to $n_r > 12$ when identifying DGE irrespective of fold-change is important. For $n_r < 12$, superior TPR makes *edgeR* the leading tool tested. For $n_r \geq 12$, minimizing false positives is more important and *DESeq* outperforms the other tools.

RNA-seq data have been submitted to ENA archive with project ID PRJEB5348.

## 1  Introduction

RNA-seq has now supplanted microarrays as the technology of choice for genome-wide Differential Gene Expression (DGE) experiments. In any experimental design, selecting the appropriate number of biological replicates is a trade-off between cost and precision. For microarray methods it has been shown that low replicate experiments often have insufficient statistical power to call DGE correctly (Pan et al. 2002) and cannot accurately measure the natural biological variability (Churchill 2002). Although it is widely appreciated that increasing the number of replicates in an RNA-seq experiment usually leads to more robust results (Auer and Doerge 2010; Hansen et al. 2011; Busby et al. 2013; Liu et al. 2014), the precise relationship between replicate number and the ability to correctly identify the differentially expressed genes (i.e., the statistical power of the experiment) has not been fully explored.

The rise of RNA-seq technology has led to the development of many tools for analyzing DGE from these data (e.g., Anders and Huber 2010; Hardcastle and Kelly 2010; Robinson et al. 2010; Wang et al. 2010; Tarazona et al. 2011; Li et al. 2012; Lund et al. 2012; Trapnell et al. 2012; Li and Tibshirani 2013; Frazee et al. 2014; Law et al. 2014;

Love et al. 2014; Moulos and Hatzis 2014). Each tool makes assumptions about the statistical properties inherent to RNA-seq data and they exploit a range of normalization and analysis techniques to compute the magnitude of a DGE result and estimate its significance. Several studies have generated data specifically for the purpose of testing the assumptions intrinsic to DGE methods (Marioni et al. 2008; Consortium 2014), but most rely either on RNA-seq datasets designed to test biological hypotheses (Bullard et al. 2010; Rapaport et al. 2013; Seyednasrollah et al. 2013) or simulated data (Busby et al. 2013; Soneson 2014), or a combination of the two (Kvam et al. 2012; Li et al. 2012; Dillies et al. 2013; Guo et al. 2013; Soneson and Delorenzi 2013; Burden et al. 2014). The majority of studies based on analysis of experimental RNA-seq data rely on data from experiments with fewer than five replicates per condition (Marioni et al. 2008; Bullard et al. 2010; Kvam et al. 2012; Li et al. 2012; Busby et al. 2013; Dillies et al. 2013; Rapaport et al. 2013; Consortium 2014; Soneson 2014), limiting their ability to compare the performance of DE tools as a function of replication.

Two studies explore higher replication by exploiting publically available RNA-seq data from 21 individual clones of two laboratory strains of mouse (Bottomly et al. 2011; Soneson and Delorenzi 2013; Burden et al. 2014). Burden et al. (2014) consider false discovery rate (FDR) as the main metric for ranking five tools and conclude that at least six replicates per condition and multiplexing DGE tools gives the best results. Soneson and Delorenzi (2013) focus on the degree of concordance between tools as a metric for comparison and conclude that none of the eleven tools they tested perform well with fewer than three replicates. Nevertheless, since the experiments are from individual mice, the data may reflect inter-individual variance in RNA expression as well as from other aspects of the experimental protocol. The same is true of studies in human that make use of data from individuals to explore higher sample replication in DGE (Guo et al. 2013; Seyednasrollah et al. 2013). Guo et al. (2013) expand replicate number by comparing six tools using RNA-seq data from breast cancer tumor-normal paired samples from fifty three individuals in The Cancer Genome Atlas (TCGA, Cancer Genome Atlas Research 2008), using this primarily to guide the construction of a simulated dataset. They conclude that all six of the tools they test suffer from oversensitivity but that *edgeR* represents the best compromise between accuracy and speed. Seyednasrollah et al. (2013) examine the performance of eight tools using mouse data (Bottomly et al. 2011) and lymphoblastoid cell data from a cohort of fifty-six unrelated Nigerian individuals from the HapMap project (International HapMap 2005). They recommend *limma* and *DESeq* for data with fewer than five replicates per condition, finding that *edgeR* is "oversensitive" and suffers from high variability in its results while *SAMSeq* suffers from a lack of statistical power with few replicates. The idea of combining DGE methods is implemented in the novel tool *PANDORA* which weights the results of different DGE tools according to their performance on test data and performs at least as well as the constituent tools (Moulos and Hatzis 2014).

In this paper the performance of DGE tools is evaluated through the first highly-replicated RNA-seq experiment designed specifically to test both the assumptions intrinsic to RNA-seq DGE tools (Gierliński et al. 2015) and to assess their performance. The paper focuses on nine popular RNA-seq specific DGE tools (as judged by citations): *baySeq*, *cuffdiff*, *DESeq*, *edgeR*, *limma*, *NOISeq*, *PoissonSeq*, *SAMSeq* and *DEGSeq* (see Table 1 for references) and assesses their performance as a function of replicate number and fold-change.

The study provides general recommendations on:

1. How many replicates future RNA-seq experiments require to maximise the sensitivity and accuracy of DGE identification and quantification.
2. The most appropriate DGE tools to use to detect DE genes in RNA-seq experiments with a given number of replicates.

**Table 1:** RNA-seq differential gene expression tools and statistical tests.

| Name | Assumed Distribution | Normalization | Description | Version | Citations[4] | Reference |
|---|---|---|---|---|---|---|
| *t-test* | normal | DEseq[1] | two-sample t-test for equal variances | - | - | - |
| *log t-test* | log-normal | DEseq[1] | log-ratio t-test | - | - | - |
| *Mann-Whitney* | none | DEseq[1] | Mann-Whitney test | - | - | Mann and Whitney (1947) |
| *Permutation* | none | DEseq[1] | permutation test | - | - | Efron and Tibshirani (1993) |
| *Bootstrap* | normal | DEseq[1] | bootstrap test | - | - | Efron and Tibshirani (1993) |
| *baySeq[3]* | negative binomial | Internal | Empirical Bayesian estimate of posterior likelihood | 1.8 | 109 | Hardcastle and Kelly (2010) |
| *Cuffdiff* | negative binomial | Internal | unknown | 2.1.1 | 481 | Trapnell et al. (2012) |
| *DEGseq[3]* | binomial | None | random sampling model using Fisher's exact test and the likelihood ratio test | 1.10.0 | 215 | Wang et al. (2010) |
| *DESeq[3]* | negative binomial | DEseq[1] | Shrinkage variance | 1.4.1 | 1204 | Anders and Huber (2010) |
| *edgeR[3]* | negative binomial | TMM[2] | Empirical Bayes estimation & an exact test analogous to Fisher's exact test but adapted to over-dispersed data | 2.2.5 | 822 | Robinson et al. (2010) |
| *Limma[3]* | Log-normal | TMM[2] | Generalised linear model | 3.4.4 | 15 | Law et al. (2014) |
| *NOISeq[3]* | None | RPKM | Non-parametric test based on signal-to-noise ratio | 29/04/2011 | 113 | Tarazona et al. (2011) |
| *PoissonSeq[3]* | Poisson log-linear model | Internal | Score statistic | 1.1 | 25 | Li et al. (2012) |
| *SAMSeq[3]* | None | Internal | Mann-Whitney test with Poisson resampling | 2.0 | 26 | Li and Tibshirani (2013) |

[1] see Anders and Huber (2010), [2] see Robinson and Oshlack (2010), [3] R (v2.15.1) & bioconductor (v2.10)
[4] as reported by PubMed Central articles that reference the listed reference (28th Jan 2015)

## 2 Tool-specific Gold Standards

RNA was sequenced from 48 biological replicate samples of *Saccharomyces cerevisiae* in each of two well-studied experimental conditions; wild-type (WT) and a *Δsnf2* mutant. Quality control and data processing steps reject several replicates from each condition resulting in 42 WT and 44 *Δsnf2* biological replicates of 'clean' data totaling ~889M aligned reads (see Materials and Methods for a full description on the experiment, the mutant strain, the sequencing and the quality control and data processing steps).

The performance of each DGE tool as a function of replicate number and expression fold-change was evaluated by comparing the DGE results from sub-sets of these replicates against the 'gold standard' set of DGE results obtained for each tool with the full set of replicates. The tool-specific gold-standards were computed by running the tool on the read-count-per-gene measurements from the full set of clean data and marking as "significantly differentially expressed" (SDE) those differentially expressed genes with multiple testing corrected *p*-values or FDRs ≤ 0.05. These gold-standard runs typically result in ≥65% of the 7,126 genes in the Ensembl v68 (Flicek et al. 2011) *S. cerevisiae* annotation being identified as SDE (except for *DEGSeq* which calls considerably more genes as SDE and *NOISeq*, which calls far fewer, see Supp. Figs. S3 & S6: A).

With the tool-specific gold-standards defined, each DGE algorithm was run iteratively on *i* repeated sub-selections drawn from the set of clean replicates (without replacement). For each of the tools, bootstrap runs were performed with $i = 100$ iterations and $n_r = 3, \dots, 40$ replicates in each condition (*cuffdiff* was significantly slower than the other tools so the number of iterations was reduced to $i = 30$ for this tool). For a given value of $n_r$, the mean log-2 transformed fold-change ($\log_2(FC)$) and median adjusted *p*-value or FDR calculated across all the bootstrap iterations was considered representative of the measured behavior for each individual gene. Again, genes were marked as SDE when the adjusted *p*-value or FDR was ≤ 0.05. From these results, true positive, true negative, false positive and false negative rates (hereafter TPR, TNR, FPR, FNR) were then calculated as a function of $n_r$ for four arbitrary fold-change thresholds ($|\log_2(FC)| = T \in \{0, 0.3, 1, 2\}$), by comparing the SDE genes from each bootstrap with the SDE genes from the tools gold-standard (see Materials and Methods for a detailed description of these calculations). Intrinsic to this method of measuring each tool's performance is the assumption that the large number of replicates in the full dataset will enable each tool to unambiguously identify the 'true' differentially expressed genes in the experiment.

## 3  Results

Figure 1 shows an example of the key performance data for *edgeR* (similar figures for the other tools can be found in Supp. Figs. S2-S9). The fraction of all genes *edgeR* calls as SDE increases as a function of $n_r$ and the impact of sampling effects on this fraction shrinks as $n_r$ increases (Figure 1: A). The TPR performance of *edgeR* changes as a function of both replicate number and fold-change threshold (Figure 1: B & C). However, *edgeR* successfully controls its FDR for all combinations of both $n_r$ and $T$ and the primary effect of increasing the number of replicates or imposing a fold-change threshold is to increase the sensitivity of the tool, converting false negatives to true positives (Figure 1: D).

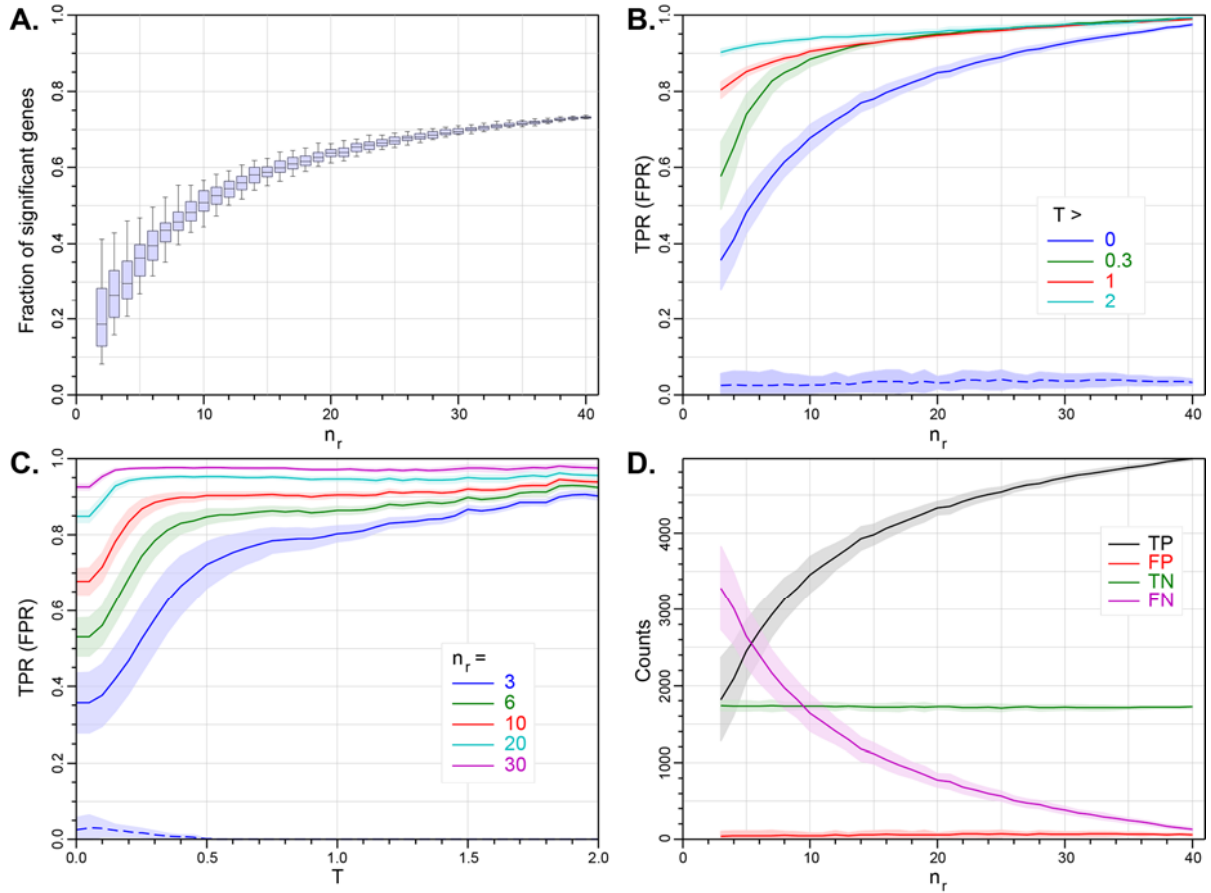

Fig 1: Statistical properties of *edgeR* as a function of $|\log_2(FC)|$ threshold, T, and the number of replicates, $n_r$. Individual data-points are not shown for clarity; however the points comprising the lines are each an average over 100 bootstrap iterations, with the shaded regions showing the 1-standard-deviation limits. A: The number of genes called as SDE as a function of the number of replicates (boxplots show the median, quartiles and 95% limits across replicate selections within a bootstrap run). B: mean true positive rate (TPR) as a function of $n_r$ for four thresholds $T \in \{0, 0.3, 1, 2\}$ (solid curves, the mean false positive rate (FPR) for $T = 0$ is shown as the dashed blue curve, for comparison). Data calculated for every $\Delta n_r = 1$. C: mean TPR as a function of $T$ for $n_r \in \{3, 6, 10, 20, 30\}$ (solid curves, again the mean FPR for $n_r = 3$ is shown as the dashed blue curve, for comparison). Data calculated every $\Delta T = 0.1$ D: The number of genes called as true/false positive/negative (TP, FP, TN and FN) as a function of $n_r$. The FPR remains extremely low with increasing $n_r$ demonstrating that *edgeR* is excellent at controlling its false discovery rate. Data calculated for every $\Delta n_r = 1$.

Figure 2 summarises the performance of all nine tools considered in this study as a function of replicate number and fold-change threshold. The TPR for bootstrap sub-selections with three replicates and no fold-change threshold ($n_r = 3, T = 0,$) is ~20-40% for all the tools except *NOISeq* and *DEGSeq*, indicating that with this few replicates these experiments were unable to identify the majority of DE genes regardless of the tool used to analyse the data (Figure 2: A).

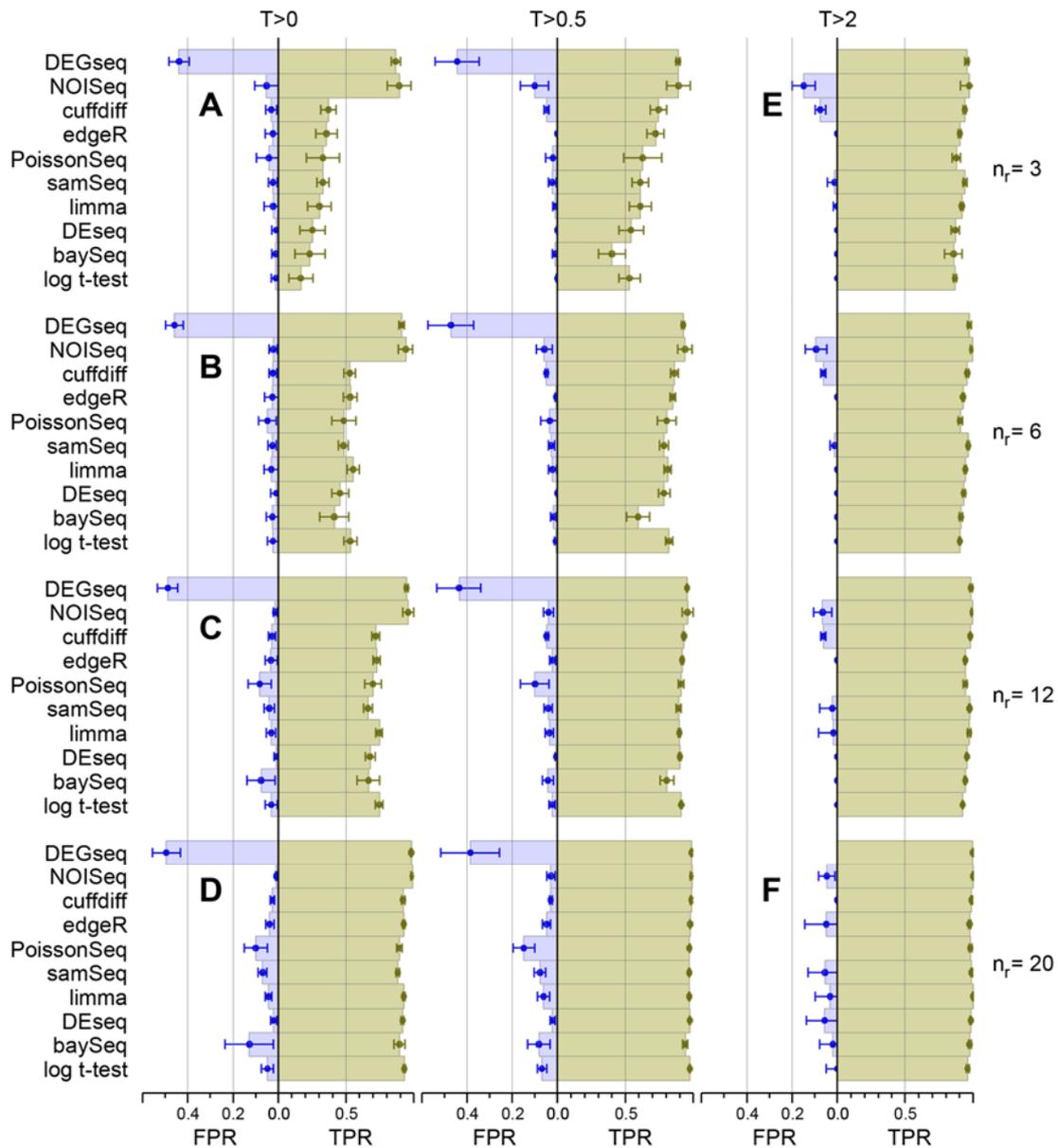

Fig 2: Comparison of the true positive rate (TPR) and false positive rate (FPR) performance for each of the DE tools on low-, medium- and highly-replicated RNA-seq data ($n_r \in \{3, 6, 12, 20\}$ – rows) for three $|\log_2(FC)|$ thresholds ($T \in \{0, 0.5, 2\}$ - columns). The TPRs & FPRs for each tool are calculated by comparing the mean number of true and false positive (TPs and FPs) calculated over 100 bootstrap iterations to the number of TPs and FPs calculated from the same tool using the full clean dataset (error-bars are 1 standard deviation). Although the TPRs and FPRs from each tool are calculated by comparing each tool against itself rather than a tool-independent 'gold standard' (albeit with the full clean dataset) the results are comparable across tools except for *DEGSeq* which calls a significantly larger fraction of genes as DE for all values of T and $n_r$ – (Supp. Fig. S4) and *NOISeq* which calls a significantly smaller fraction of genes as DE for all values of T and $n_r$ – (Supp. Fig. S7). *DEGSeq* and *PoissonSeq* produce no FPs for the highest threshold (T>2) and thus no FPR is shown for them. In general, the FPR decreases with increasing fold-change threshold, and for the highest fold-change genes the FPRs are formally consistent with zero. There are exceptions however; *PoissonSeq* & *BaySeq* show an increasing FPR with increasing $n_r$ and *cuffdiff* unexpectedly shows an increase in FPR with increasing fold-change threshold. *DESeq* appears more conservative than the other tools, consistently returning fewer FPs (particularly for high values of $n_r$) and fewer TPs (particularly at low values of $n_r$).

*DEGSeq* shows strong TPR performance but this is coupled with a high FPR due to *DEGSeq* overestimating the number of SDE genes regardless of the number of replicates (Supp. Fig. S4: A). *NOISeq* shows a similarly strong TPR performance with an apparently low FDR as well, however *NOISeq* identifies only a few hundred of the most strongly DE genes as SDE regardless of the number of replicates, demonstrating that this tool is insensitive to the majority of DE (Supp. Fig. S7: A). Excluding *DEGSeq* and *NOISeq*, the TPR performance for all the remaining tools is a function of fold-change (Figure 1: C & Supp. Figs. S2-S9: C). For the highest fold-change genes ($T = 2$), all the remaining seven tools show TPRs $\gtrsim$ 85% and (with the exception of *cuffdiff*) FPRs consistent with zero (Figure 2: E). These tools are successfully capturing the majority of the true differential expression signal for the most strongly changing genes from each tools gold standard with as few as three replicates per condition. For this cohort of high fold-change SDE genes the TPR is largely insensitive to replicate number. Irrespective of the tool, increasing the number of replicates to $n_r = 20$ for $T = 2$ provides only a modest increase in TPR from ~85% to ~95% (Figure 2: F, Figure 1: B & Supp. Figs. S2-S9: B). Increasing the number of replicates has a dramatic effect on the detection rate of genes with smaller fold-changes. Achieving an ~85% detection rate with *edgeR* for fold-change thresholds of $T = 1$, 0.3 & 0 requires ~9, 11 & 26 replicates, respectively (Figure 1: B & C). This behavior is reflected across all seven tools (Figure 2: A-D). Reducing the fold-change threshold reduces the TPR independently of replicate number for all the tools except *DEGSeq*, *NOISeq* and *baySeq*. The TPR performance as a function of fold-change threshold has two distinct linear regions: a shallow linear regime at high-$T$ and a steeper region at low-$T$ (Figure 1: C & Supp. Figs. S2-9: C). The transition between these two regions is a function of both the tool and the number of replicates. For *edgeR* with $n_r = 3$, this transition fold-change threshold is ~0.5 and drops to ~0.25 and ~0.15 for $n_r = 10$ & 30 respectively (Figure 1: C). These transitions represent an optimal fold-change threshold to filter the data by to maximise both the quality and the utility of the data.

The best performing tools *edgeR*, *DESeq* and *limma* successfully control their FPR, maintaining it consistently close to, or below 5% irrespective of fold-change threshold or number of replicates (Figure 1: B & C & Supp. Figs. S5 & S6: B & C) highlighting again that the primary effect of increasing replicate number is to increase the sensitivity of these tools, converting false negatives to true positives (Figure 1: D & Supp. Figs. S5 & S6: D). Other tools are not so successful in this regard but a detailed interpretation of the FPR from this test is complicated by the fact that each tool is tested against its own gold standard. A more robust method for probing the FPR performance of DE tools is presented below.

## 4 Tool consistency with high replicate data

The DGE tool performance tests described here assume that, given enough replicates, the tools converge on the true underlying differential expression signal in the data. This assumption was tested by clustering the DE measurements for each tool's 'gold standard' along with the results from five additional simple statistical tests applied to the same data (see Materials and Methods for a detailed description of the statistical tests). The lists of SDE genes from each tool and test (adjusted p-value or FDR threshold $\leq 0.05$) were hierarchically clustered with the R package *pvclust* (Figure 3, Suzuki and Shimodaira 2006). *pvclust* uses bootstrapping to compute the statistical significance of sub-clusters within the dendrogram. Approximately Unbiased p-value percentages (AU% – Figure 3, bracketed values) calculated for each branch in the clustering are an indication of how robust each branch is to sampling error. The most widely used tools (*edgeR* and *limma*, Table 1) are tightly grouped in a robust cluster with the standard statistical tests (Figure 3, cluster 1). In 90% of the bootstraps, *DESeq* also clusters tightly with this group, and when it does, *cuffdiff* also clusters robustly with this group, suggesting these tools and tests are reliably and reproducibly converging on approximately the same answer, given a large number of replicates (Figure 3, cluster 2). Several of the standard statistical tests are non-parametric (Mann-Whitney, permutation and bootstrap) and use very different underlying methods compared to the tools in this cluster, indicating that the agreement of techniques within this group is not the result of a similar underlying methodology, but is likely reflective of the true differential expression signal in the data. The remaining outlier tools (*PoissonSeq*, SAMSeq, *NOISeq*, *DEGSeq* and *baySEQ*) cluster independently suggesting that these tools reach a different result to those in Cluster 1. Furthermore, the outlier tools cluster into three robust groups (Figure 3, clusters 3-5), indicating that they are reaching different conclusions to each other as well as to the non-outlier tools and are unlikely to be capturing the true underlying differential expression signal in the data.

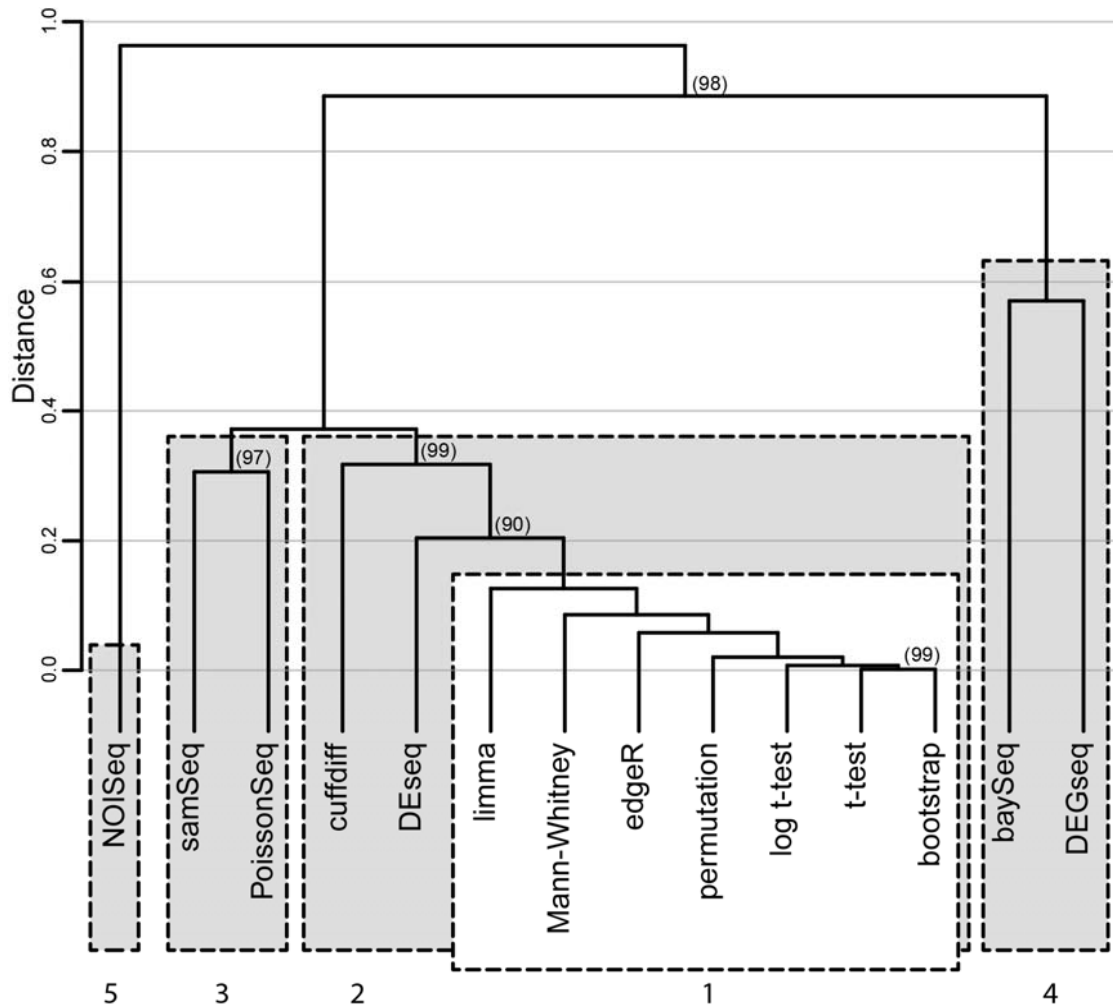

Fig 3: Hierarchical clustering of nine RNA-seq DE tools and five standard statistical tests using all the full clean dataset comprising 42 WT & 44 *Δsnf2* replicates. Lists of significantly differentially expressed (SDE) genes from each tool and test (adjusted p-value or FDR threshold ≤ 0.05) were hierarchically clustered with the R package *pvclust* (Euclidian distance metric, maximum-linkage) de. Approximately Unbiased p-value percentages (bracketed values) calculated for each branch in the clustering represent the support in the data for the observed sub-tree clustering. AU%>95% are strongly supported by the data. AU% values are not shown for branch-points where AU%=100 for clarity. The outlier clustering of *NOISeq*, *baySeq*, *DEGSeq*, *SAMSeq*, and *PoissonSeq* suggest that these tools are clearly distinct from the other tools. The 90% AU value for *DESeq* indicates that this tool clusters tightly with Cluster 1 90% of the time, and that when it does *cuffdiff* also clusters robustly with this group. Combined with the tool performance data shown in Figure 3, this suggests that, given a large number of replicates, the tools and tests in Cluster 1 (and to a lesser extent in Cluster 2) are reliably and reproducibly converging on a similar answer, and are likely to be correctly capturing the SDE signal in the data.

## 5 Testing Tool False Positive Rates

Perhaps the most important performance measure for RNA-seq DE tools is their false detection rate. The large number of replicates in this study permits a simple test of the FPR for each of the tools. Two sets of $n_r$ replicates were randomly selected (without replacement) from the WT condition. Under the null hypothesis that there is no expression change between these two sets, every gene identified as SDE is, by definition, a false positive. For each bootstrap run, the fraction of the total gene set identified as SDE was computed. The distribution of this false positive fraction as a function of the number of replicates, for each DE tool, is shown in Figure 4. This approach shows that *DEGSeq* and *SAMSeq* perform poorly even with a large number of replicates. *DEGSeq*, in particular, has poor FP performance with every bootstrap iteration identifying >5% of all genes as false positives (FPs) and a median FPR of ~50% irrespective of the number of replicates. Approximately 10% of *PoissonSeq* and *cuffdiff*, and 40% of *SAMSeq* bootstrap iterations identify >5% of all genes as FPs, suggesting that although better than *DEGSeq*, these tools are not controlling their FDR well. These results indicate that *PoissonSeq, cuffdiff*, *DEGSeq* and *SAMSeq* have inferior performance to the other tools considered in this study for RNA-seq DE analysis.

## 6 Discussion

In this work the performance of nine popular RNA-seq DE tools has been evaluated using a highly-replicated two-condition RNA-seq experiment designed specifically for the purpose of benchmarking RNA-seq DGE tools on genuine biological replicate data. Three of the nine tools, *edgeR, limma* and *DESeq* show excellent performance in the tests presented here. With the exception of *limma*, these are the most widely-used of the tools tested here as measured by citations (Table 1), suggesting that the majority of the RNA-seq DE analyses in the literature are using the most appropriate tools for the job. An additional important feature of all three tools is that they allow confounding experimental factors to be specified for DGE analysis. This permits *edgeR*, *limma* and *DESeq* to be used even with challenging datasets.

For experiments where it is important to capture as many of the truly SDE genes as possible but with a low number of replicates (*i.e.*, $n \lesssim 12$), the data presented here suggest *edgeR* in preference to *limma* or *DESeq* due to its superior TP identification rate. For experiments with sufficient numbers of replicates to ensure that the majority of the true SDE is already being captured (*i.e.*, $n \gtrsim 12$), and where it is, instead, important to minimise the number of false positives, the slightly better FPR performance of *DESeq* suggests it should be the tool of choice. Conversely, *PoissonSeq, SAMSeq, NOISeq, DEGSeq, baySeq* and *cuffdiff* all show inferior performance compared to *edgeR*, *DESeq* and *limma*. Table 2 summarises recommendations for choosing RNA-seq DE tools, based on the results of these benchmarking tests. It is clear from the benchmarking runs that even the best tools have limited statistical power with few replicates in each condition, unless a stringent fold-change threshold is imposed (Figure 2). For all the tools except *baySeq*, however, the FPR, is approximately constant regardless of replicate number (Figure 4) suggesting that controlling the FNR rather than the FPR is the primary justification for imposing such a fold-change threshold. Interestingly, the variation intrinsic to the experimental procedures and protocols will result in a hard lower limit on the detectable fold-changes for biologically relevant DE. Unfortunately, it is not possible to calculate this limit here using the gene count data alone since it requires prior knowledge of actual fold-changes to measure the impact of experimental variance.

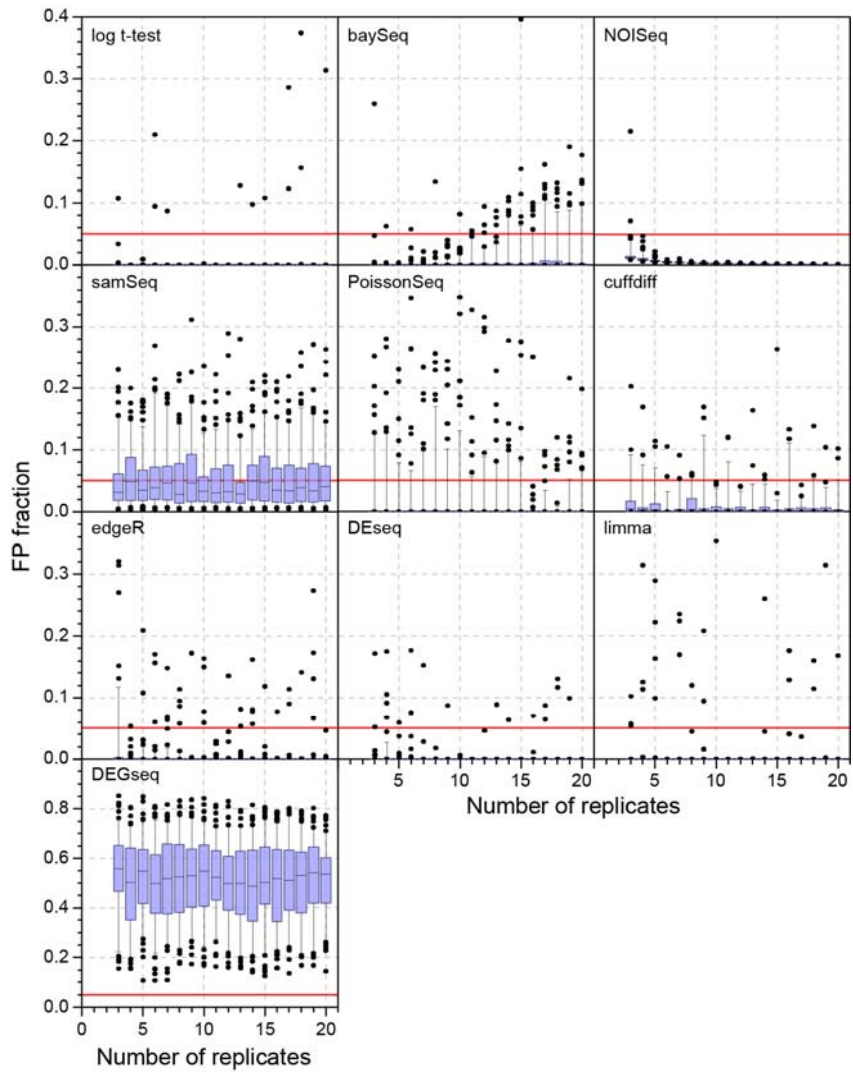

Fig 4: Testing false positive rate (FPR) performance: Each tool was used to call significantly differentially expressed (SDE) genes based on two artificial 'conditions' each constructed only from WT biological replicates. Genes identified as SDE are, by definition, false positives. The box plots show the median, quartiles and 95% data limits on the FPR for 100 bootstrap iterations of each of the nine tools and the log t-test for $n_r = 3, 4, .., 20$. The red line highlights a 5% FDR. In most cases the tools perform well for each bootstrap iteration, with only a small number of iterations showing a FPR >5%. Only *DEGSeq*, here shown with a different y-axis scale to the rest of the plots, and to a lesser extent *SAMSeq*, consistently show a high FPR, suggesting that they are struggling to control their FDR adequately.

When designing and RNA-seq experiment with the primary goal of identifying those SDE genes that change by more than a factor of two ($T = 1$), three clean replicates per condition may be sufficient. However, this is not the same as conducting the experiment with a total of three replicates, because there is a significant minority chance that one-or-more replicate within each condition should be rejected (see Gierliński et al. 2015). Conversely, for biological questions in which identifying the majority of the DE genes is important, a low-replicate experiment may not provide a sufficiently detailed view of the differential expression to inform the biology accurately. In these situations, it would be prudent to obtain at least 12 clean replicates per condition allowing the identification of ≳90% of the truly SDE genes with $T \gtrsim 0.3$ by any of the tools presented here except *NOISeq* or *DESeq*. ). It is worth recalling that identifying a gene as SDE does not necessarily equate to identifying it as biological significant and that it is important to consider both the magnitude of the measured fold-change and existing biological knowledge alongside the statistical significance when inferring a biological significance for the results of DGE experiments.

The experiment performed here is likely to be a best-case scenario and so represent the upper limit in performance of the tools tested. *S. cerevisiae* is one of the best-studied model organisms in biology, with a genome that is relatively small and well understood, few genes containing more than a single exon and there is no tissue-specific gene expression. In an organism with a more complex transcriptome, the performance of all the DE tools is likely to be worse due to the presence of multiple transcript isoforms, anti-sense non-coding RNA transcription and incomplete or poorly known annotations, particularly for 5' and 3' UTRs (Schurch et al. 2014). Although the majority of current DGE tools including the nine analysed here, rely on an existing genome annotation, the recently published DE tool *derfinder* (Frazee et al. 2014) examines DE for any region of a genome without annotations by analysing DE at base-pair resolution and grouping adjacent regions with similar DE signals. Such annotation-free differential expression tools may well represent the future for differential expression studies with RNA-seq data since they have the potential to mitigate the impact of genome annotation on detection of DE.

To the best of our knowledge the experiment presented here is the most highly replicated RNA-seq dataset to date and the only one specifically designed for testing the process of calling differential expression. As such, it will be a useful resource for the bioinformatics community as a test-bed for tool development, and for the wider biological science community as the most detailed description of transcription in wild-type & *Δsnf2* mutant *S. cerevisiae*. All the code for this work will be publically available (Github), and both the raw and processed data can be accessed from the European Nucleotide Archive (project ID: PRJEB5348).

**Table 2:** A Summary of the recommendations of this paper

|  | Agreement with other tools[1] | WT v WT FPR[2] | Fold-change Threshold (T)[3] | Tool recommended for: (# good replicates per condition)[4] | | |
|---|---|---|---|---|---|---|
|  |  |  |  | <=3 | <=12 | >12 |
| *BaySeq* | inconsistent | Pass |  |  |  |  |
| *cuffdiff* | consistent | Fail |  |  |  |  |
| *DEGSeq* | inconsistent | Fail |  |  |  |  |
| **DESeq** | consistent | Pass | 0 |  |  | Yes |
|  |  |  | 0.5 |  | Yes | Yes |
|  |  |  | 2 | Yes | Yes | Yes |
| **edgeR** | consistent | Pass | 0 |  |  | Yes |
|  |  |  | 0.5 | Yes | Yes | Yes |
|  |  |  | 2 | Yes | Yes | Yes |
| **Limma** | consistent | Pass | 0 |  |  | Yes |
|  |  |  | 0.5 |  | Yes | Yes |
|  |  |  | 2 | Yes | Yes | Yes |
| *NOISeq* | inconsistent | Pass |  |  |  |  |
| *PoissonSeq* | inconsistent | Fail |  |  |  |  |
| *SAMSeq* | inconsistent | Fail |  |  |  |  |

[1] Full clean replicate dataset, see 'Tool consistency with high replicate data' and Figure 3.
[2] See 'Testing Tool False Positive Rates' and Figure 4
[3] See 'Differential Expression Tool Performance as a function of replicate number'
[4] See Figure 2

## 7   Recommendations for RNA-seq experiment design

The results of this study suggest the following should be considered when designing an RNA-seq experiment for DGE:

1) At least 6 replicates per condition for all experiments.
2) At least 12 replicates per condition for experiments where identifying the majority of all DE genes is important.
3) For experiments with < 12 replicates per condition; use *edgeR*.
4) For experiments with > 12 replicates per condition; use *DESeq*.

5) Apply a fold-change threshold appropriate to the number of replicates per condition between $0.1 \leq T \leq 0.5$ (see Figure 2 and the discussion of tool performance as a function of replication).

# 8 Materials & Methods

## 8.1. *Δsnf2* mutant

*Saccharomyces cerevisiae* is one of the best studied organisms in molecular biology with a relatively small transcriptome and very limited alternative splicing and was chosen in order to give us the simplest RNA-seq data possible. *SNF2* is the catalytic subunit of ATP dependent chromatin remodeling SWI/SNF complex in yeast. *SNF2* forms part of a transcriptional activator and mutation in *SNF2* brings about significant changes in transcription (e.g., Neigeborn and Carlson 1984; Stern et al. 1984; Peterson et al. 1991; Hirschhorn et al. 1992; Peterson and Herskowitz 1992; Holstege et al. 1998; Sudarsanam et al. 2000; Becker and Horz 2002; Gkikopoulos et al. 2011; Ryan and Owen-Hughes 2011, and references therein).

## 8.2. S. cerevisiae growth conditions & RNA extraction

The *S. cerevisiae* strains used in the experiment were wildtype (BY4741 strain, WT) and *Δsnf2* mutant in the same genetic background. Asynchronous WT & *Δsnf2* mutant strains were streaked out on rich media (YPAD) to get individual colonies. For 48 replicates in both strains, single colonies were inoculated to 15 ml cultures and cells were grown to an OD600 of 0.7-0.8 (corresponding to approximately $10^6$ cells) at 30°C. RNA was isolated using RNeasy mini kit (Qiagen, Manchester, UK) protocol that uses Zymolyase for yeast cell lysis and DNase treatment to remove DNA contamination. The amount of total RNA extracted ranged from 30.3 μg to 126.9 μg per sample. Although the amount RNA extracted was variable, the distributions were consistent with being drawn from the same population (Kolmogorov–Smirnov test p = 0.16) indicating no bias in RNA content between WT and *Δsnf2* mutant samples.

## 8.3. Library preparation, spike-in addition & sequencing

The RNA-seq experiment described here implements a 'balanced block design' in order to control for technical artifacts such as library batch effects (Kaisers et al. 2014), barcoding biases and lane effects via randomization of the libraries (Colbourn and Dinitz 2007; Auer and Doerge 2010). Additionally, all the replicates include artificial RNA spike-in controls in order to allow external calibration of the RNA concentrations in each sample and of the measured fold-changes between the two conditions (Jiang et al. 2011; Loven et al. 2012). The 96 samples were prepared in batches of 24 samples with 12 of each strain in each batch. Barcodes were pre-assigned randomly between the samples with barcode IDs 1-48 assigned to the *Δsnf2* mutant samples the 49-96 to the WT strain. For each batch the Illumina TruSeq protocol was used to prepare the sequencing library, with the addition of the ERCC spike-in standard (Ambion, Paisley, UK, Jiang et al. 2011). Briefly, samples were polyA enriched with poly-dT beads and 1 μl of 1:100 spike-in added to 19.5 μl of polyA enriched samples. Spike-in mix 1 was used with the *Δsnf2* mutant and mix 2 with WT. The RNA was then fragmented and subsequently underwent both first and second strand cDNA synthesis. The cDNA then was then subjected to end repair, 3' end adenylatation, and barcode sequences were added. Finally, the un-barcoded adapters were ligated, templates purified and finally the samples were enriched via barcode-specific PCR primers. At this point the quality of the libraries was examined and passed before being diluted down to 10 nM and quantified (using fluorescence-based quantification) for accurate aliquoting for cluster generation and appropriate lane loading. Seven independent pools of the 96 barcoded samples were prepared and loaded onto seven lanes of an Illumina HiSeq 2000. Thus, each lane contains all 96 samples prepared in four batches with different spike-in mixes in each strain. The flow-cell was run for 51 cycles single-end.

## 8.4. Read alignment & and read-count-per-gene measurement

The lane data were de-multiplexed and processed through Cassava pipeline v1.8 to generate 672 fastq files comprising 7 technical replicates for each of the 96 biological replicates in the experiment. A total of $\sim 10^9$ reads were reported with each technical replicate having between $0.8\text{-}2.8 \times 10^6$ reads. Aggregating the technical replicates across lanes results in $\sim 10^7$ reads per biological replicate. First pass quality control of the reads was performed with *fastQC* (http://www.bioinformatics.bbsrc.ac.uk/projects/fastqc) for each technical replicate. The reads from each technical replicate were then aligned to the Ensembl release 68 (Flicek et al. 2011) *S. cerevisiae* genome with *bowtie2* (*v2.0.0-beta7*) (Trapnell and Salzberg 2009) and *TopHat2* (*v2.0.5*) (Trapnell et al. 2009) using the following parameters: *--max-intron-length 1000 –min-intron-length 10 –microexon-search –b2-very-sensitive –max-multihits 1*. The aligned reads were then aggregated with *htseq-count* (v0.5.3p9, Anders et al. 2014) using the Ensembl v68 *S. cerevisiae* genome annotation to give total gene read counts for all 7,126 gene features for each technical replicate. Finally, the read-count-per-gene measurements for each technical replicate were summed across sequencing lanes to give read-count-per-gene for each of the 96 biological replicates and these are then used to identify poorly correlating 'bad' replicates within the two conditions that were then subsequently removed from the analysis (see Gierliński et al. 2015 for a detailed description of this process). This resulted in a total of 42 WT & 44 *Δsnf2* biological replicates of 'clean' read-count-per-gene data.

## 8.5. Tool Details and Considerations for Differential Expression Calculations

Most of the DGE tools assessed here calculate both a fold-change (typically expressed as a logarithm to base 2 of the expression ratio, $\log_2 FC$) and a statistical significance of differential expression for each gene. The fold-change is based on the mean count across replicates in each condition and for many of the tools this includes a calculation of sample-specific normalization factors based on the gene read-count data. For this study the default normalization factors were used for each of the tools assessed. Whilst there are differences between the normalizations used by these tools it has been suggested that the details of which method is used to normalise the data does not significantly alter the downstream DGE results (Seyednasrollah et al. 2013). These normalization methods do, however, rely on the assumption that the majority of genes do not change their expression levels between conditions (e.g., Dillies et al. 2013). If this assumption is not satisfied the measurements of both DGE fold-change and significance are likely to be incorrect.

The statistical significances calculated by DGE tools are usually based on the null hypothesis of no expression change between the conditions. Calculating this significance typically relies on two key factors: 1) an assumption about the probability distribution that underlies the raw read-count measurements, and 2) being able to accurately measure the mean count and variance for each gene. Different tools assume different forms for the underlying read-count distribution including the negative binomial (*baySeq*, *Cuffdiff*, *DESeq* and *edgeR*), beta-binomial (*BBSeq*), binomial (*DEGSeq*), Poisson (*PoissonSeq*), and log-normal (*limma*) distributions. A few algorithms make no assumptions about the read-count distribution and instead take non-parametric approaches to testing for DGE (*NOISeq* & *SAMSeq*). Gierliński et al. (2015) show that for this data the majority of gene expression is consistent with both log-normal and negative binomial distributions except for the lowest expression genes, for which only the negative binomial distribution remains consistent with the data. For experiments with high numbers of replicates per condition ($n \gtrsim 12$) the mean and variance estimators can be accurately computed directly on the data, However, many RNA-seq DGE studies rely on a low numbers of replicates per condition ($n \lesssim 3$) so several of the DGE tools (e.g., *edgeR*, *limma*) compensate for the lack of replication by modelling the mean-variance relation and borrowing information across genes to shrink the given gene's variance towards the common model (Cui et al. 2005; De Hertogh et al. 2010; Robinson et al. 2010). The stabilised variance helps avoid some of the spurious false positives and negatives, but is strongly dependent on an assumed read count distribution and on the assumption that the large majority of the gene counts are not truly differentially expressed. Given these methods dependence on accurate mean and variance measurements it is somewhat surprising that scientists would contemplate doing DGE analysis without replicated data, but for completeness we note that several DGE analysis tools advertise that they can work with a single replicate per condition (Anders and Huber 2010; Robinson et al. 2010; Tarazona et al. 2011).

## 8.6. Bootstrap Differential Expression Calculations

A utility pipeline was written to automate the process of running each DGE algorithm iteratively on $i$ repeated sub-selections of clean replicates. Each sub-selection is comprised of $n_r$ replicates chosen at random without replacement (that is, an individual replicate can appear only once within each sub-selection). This bootstrapping procedure includes applying the default normalization for each tool where relevant and possible (see Tool Details and Considerations for Differential Expression Calculations) and the full output for each tool was stored in a local *sqlite* database, including the log-2 transformed fold-change and the statistical significance for every expressed gene in the *S. cerevisiae* annotation. Most of the tools return Benjamini-Hochberg (hereafter BH; Benjamini and Hochberg 1995) corrected *p*-values or FDRs as their measure of statistical significance. Genes with adjusted *p*-value or FDR $\leq 0.05$ <= 0.05 were marked as "significantly differentially expressed" (SDE). Supplementary Figure S1 shows an example of the output mean $\log_2 FC$ and median p-value data for the tool *edgeR* with $n_r = 3$.

From these data, TPRs, TNRs, FPRs and FNRs for each tool were computed as a function of the number of replicates, $n_r$, for four arbitrary absolute log-2 fold-change thresholds, $T \in \{0, 0.3, 1, 2\}$. A reference fold-change was used for deciding whether each gene falls above the threshold $T$ because the measured values of mean $|\log_2 FC|$ calculated for a gene varies considerably with both the tool being used and $n_r$. These reference fold-changes were defined independently of the tools by applying *DESeq* normalization (Anders and Huber 2010) to the read-count-per-gene data from the full clean set of biological replicates for each condition and then taking the log-2 transformed ratio of the mean normalized read-count-per-gene for each condition. For each individual DE calculation within a bootstrap run (i.e., an individual DE calculation with a specific tool with a given $n_r$), each gene was called as true/false positive/negative by comparing whether it was called as SDE in the bootstrap run, and whether it was called as SDE in the corresponding tool-specific 'gold standard'. Then, taking each fold-change threshold in turn, the mean of number of true/false positives/negatives for genes with reference fold-changes above this threshold was calculated across all the individual DE calculations within a bootstrap run. This results in a TPR, TNR, FPR and FNR for a tool, for a given $n_r$ and for a given $T$ (Equations 1-4)

$$\text{TPR}(n_r, T) = \frac{\text{TP}(n_r, T)}{\text{TP}(n_r, T) + \text{FN}(n_r, T)} \tag{1}$$

$$\text{FPR}(n_r, T) = \frac{\text{FP}(n_r, T)}{\text{FP}(n_r, T) + \text{TN}(n_r, T)} \tag{2}$$

$$\text{TNR}(n_r, T) = \frac{\text{TN}(n_r, T)}{\text{TN}(n_r, T) + \text{FP}(n_r, T)} \tag{3}$$

$$\text{FNR}(n_r, T) = \frac{\text{FN}(n_r, T)}{\text{FN}(n_r, T) + \text{TP}(n_r, T)} \tag{4}$$

Uncertainties in the resulting values were calculated by propagating the standard deviations of the numbers of TPs, TNs, FPs and FNs across the calculations within each bootstrap run, to reflect the spread of calculated values due to the random sampling of replicates.

## 8.7. Standard statistical tests for differential expression

When assessing the performance of each DE tool on the full set of clean data we compare the tools not only within themselves, but also to the following set of standard statistical tests. For the following mathematical descriptions, $\boldsymbol{x}_{gk} = (x_{g1k}, x_{g2k}, \ldots, x_{gn_kk})$ is a vector of $n_k$ (clean) replicates for gene $g$ and condition $k$, $\bar{x}_{gk}$ and $s^2_{gk}$ are the mean and variance of this vector.

### 8.7.1. t-test

The null hypothesis in t-test is that the given gene under two conditions have the same mean count, $H_0: \mu_{g1} = \mu_{g2}$. We used the test statistic

$$t_g = \frac{\bar{x}_{g1} - \bar{x}_{g2}}{\sqrt{s^2_{g12}\left(\frac{1}{n_1} + \frac{1}{n_2}\right)}} \tag{5}$$

with common variance estimator $s^2_{g12} = [(n_1 - 1)s^2_{g1} + (n_2 - 1)s^2_{g2}]/\nu$ and the number of degrees of freedom is $\nu = n_1 + n_2 - 2$.

### 8.7.2. Log-ratio t-test

This modified t-test is more appropriate for log-normally distributed data. The null hypothesis is $\ln \mu_{g1} = \ln \mu_{g2}$. The test statistic,

$$t_g = \frac{\ln \bar{x}_{g1} - \ln \bar{x}_{g2}}{\sqrt{\frac{s_{g1}^2}{n_1 \bar{x}_{g1}^2} + \frac{s_{g2}^2}{n_2 \bar{x}_{g2}^2}}} \qquad (6)$$

is approximately distributed with t-distribution with $n_1 + n_2 - 2$ degrees of freedom (see Olsson 2005).

### 8.7.3. Mann-Whitney test

Mann-Whitney (Mann and Whitney (1947) - hereafter MW) test is a non-parametric test assessing if count rate in a gene under one condition tends to be larger than under the other. The null hypothesis is $H_0\colon \Pr(x_{gi1} > x_{gj2}) = \frac{1}{2}$, for each pair of replicates $i$ and $j$. p-values were calculated using normal approximation (Bellera et al. 2010) and taking ties into account (Harbord (2004), p.428-431). The MW test relies on ranks, not actual data values, which makes it distribution-free. On the other hand, when every replicate in one condition is larger than every replicate in the other condition, the MW test will return the same p-value, regardless of how much the two conditions differ.

### 8.7.4. Permutation test

In the permutation test, counts from both conditions are pooled together (for each gene), $\boldsymbol{x}_g = (\boldsymbol{x}_{g1}, \boldsymbol{x}_{g2})$ and then randomly resampled $B$ times without replacement from $\boldsymbol{x}_g$, using the original sizes, $n_1$ and $n_2$. For the $b$-th random permutation $\boldsymbol{x}_{g1}^*(b)$ and $\boldsymbol{x}_{g2}^*(b)$ we find the test statistic, $D_g^*(b) = \bar{x}_{g1}^*(b) - \bar{x}_{g2}^*(b)$, which is the difference between the means of the two sampled vectors. This is compared with the observed statistic $D_g = \bar{x}_{g1} - \bar{x}_{g2}$. The test p-value is the fraction of cases where the resampled statistic exceeds the observed one, $p_g = |D_g^*(b)| > |D_g|/B$ (for more details see Efron and Tibshirani (1993), p203-219). The advantage of the permutation test is that it does not make any assumptions about the underlying distribution, but rather models it directly from data. The disadvantage is that it requires many replicates to build this underlying distribution, as it is not applicable for a typical experiment with, say, three replicates.

### 8.7.5. Bootstrap test

The Studentised bootstrap test described by Efron and Tibshirani (1993, p220-224) was employed here. It estimates probability distribution of the two populations under the null hypothesis of the common mean. Data are resampled with replacement to estimate the significance level. For the b-the bootstrap, $\boldsymbol{x}_{g1}^*(b)$ and $\boldsymbol{x}_{g2}^*(b)$, the test statistic is

$$t_g^*(b) = \frac{\bar{x}_{g1}^*(b) - \bar{x}_{g2}^*(b)}{\sqrt{s_{g12}^{*2}(b)\left(\frac{1}{n_1} + \frac{1}{n_2}\right)}} \qquad (7)$$

where the common variance estimator is $s_{g12}^{*2}(b) = [(n_1 - 1)s_{g1}^{*2}(b) + (n_2 - 1)s_{g2}^{*2}(b)]/(n_1 + n_2 - 2)$. This is compared with the observed statistic (Equation 5). As in the permutation test, the test p-value is the fraction of cases where the resampled statistic exceeds the observed one, $p_g = |t_g^*(b)| > |t_g|/B$.

## 9 Acknowledgements

We acknowledge Dr Tom Walsh for his efforts and support in managing the software requirements for this project on our HPC platform and thank Dr Gabriele Schweikert for stimulating discussions. This work was supported by: The Wellcome Trust strategic awards WT09230, WT083481 & WT097945, Biotechnology and Biological Sciences Research Council (BBSRC - BB/H002286/1 & BB/J00247X/1). Additional funding for Cole, Gierliński, and Schofield - The Wellcome Trust (92530/Z/10/Z); Schurch – BBSRC grant BB/M004155/1; Sherstnev - BBSRC grant BB/H002286/1; Blaxter, Wrobel & Gharbi (GenePool) - MRC (MR/K001744/1); Owen-Hughes & Singh – The Wellcome Trust senior fellowship 095062; Prof. Simpson's lab is funded by The Scottish Government and BBSRC grants BB/M004155/1 and BB/H002286/1.

# 11 Supplementary Figures

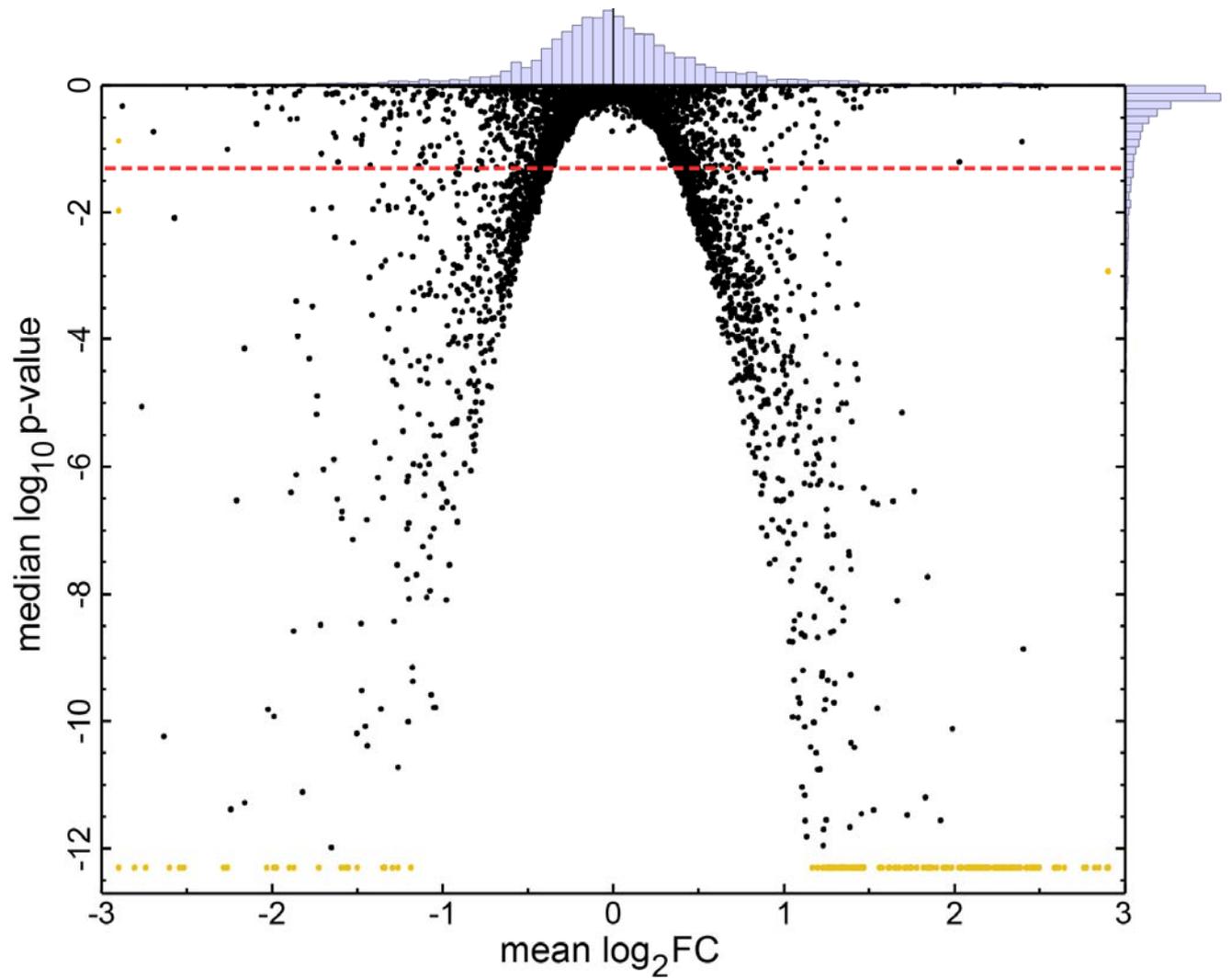

Fig S1: An example of the log₁₀(BH-corrected p-value) vs $log_2(FC)$ output data from the differential expression algorithms in this study. The data shown here are from the tool *edgeR*, averaged over 100 bootstrap iterations with $n_r = 3$. *edgeR* outputs fold-changes and p-values for 6,885 (96.6%) of the 7,126 input genes, 1,571 (22.1%) of which have BH-corrected p-values $\leq 0.05$ (dashed red line). Points shown in yellow represent data with values outside the axes limits.

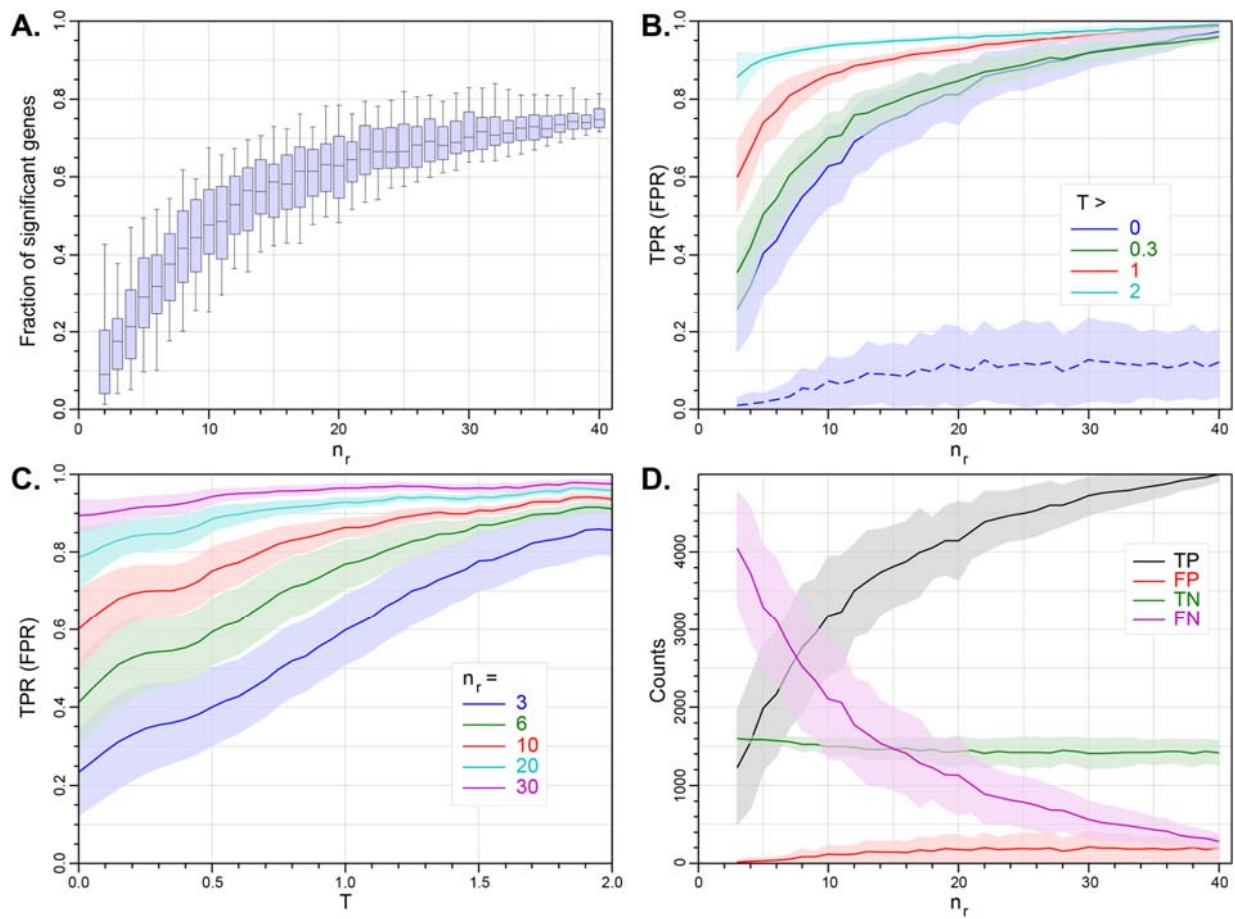

Fig S2: Statistical properties of *BaySeq* as a function of $|\log_2 FC|$ threshold, T and the number of replicates, $n_r$. As in Figure 2, individual data-points are not shown for clarity; however the points comprising the lines are each an average over 100 bootstrap iterations, with the shaded regions showing the 1-standard-deviation limits. A: The number of genes called as SDE as a function of the number of replicates (boxplots show the median, quartiles and 95% data limits). B: mean TPR as a function of $n_r$ for four fold-change thresholds $T \in \{0, 0.3, 1, 2\}$ (solid curves, the mean FPR for $T = 0$ is shown as the dashed blue curve, for comparison). Data calculated every $\Delta n_r = 1$. C: mean TPR as a function of T for $n_r \in \{3, 6, 10, 20, 30\}$ (solid curves, again the mean FPR for $n_r = 3$ is shown as the dashed blue curve, for comparison). Data calculated every $\Delta T = 0.1$ D: The number of genes called as TP, FP, TN and FN as a function of $n_r$.

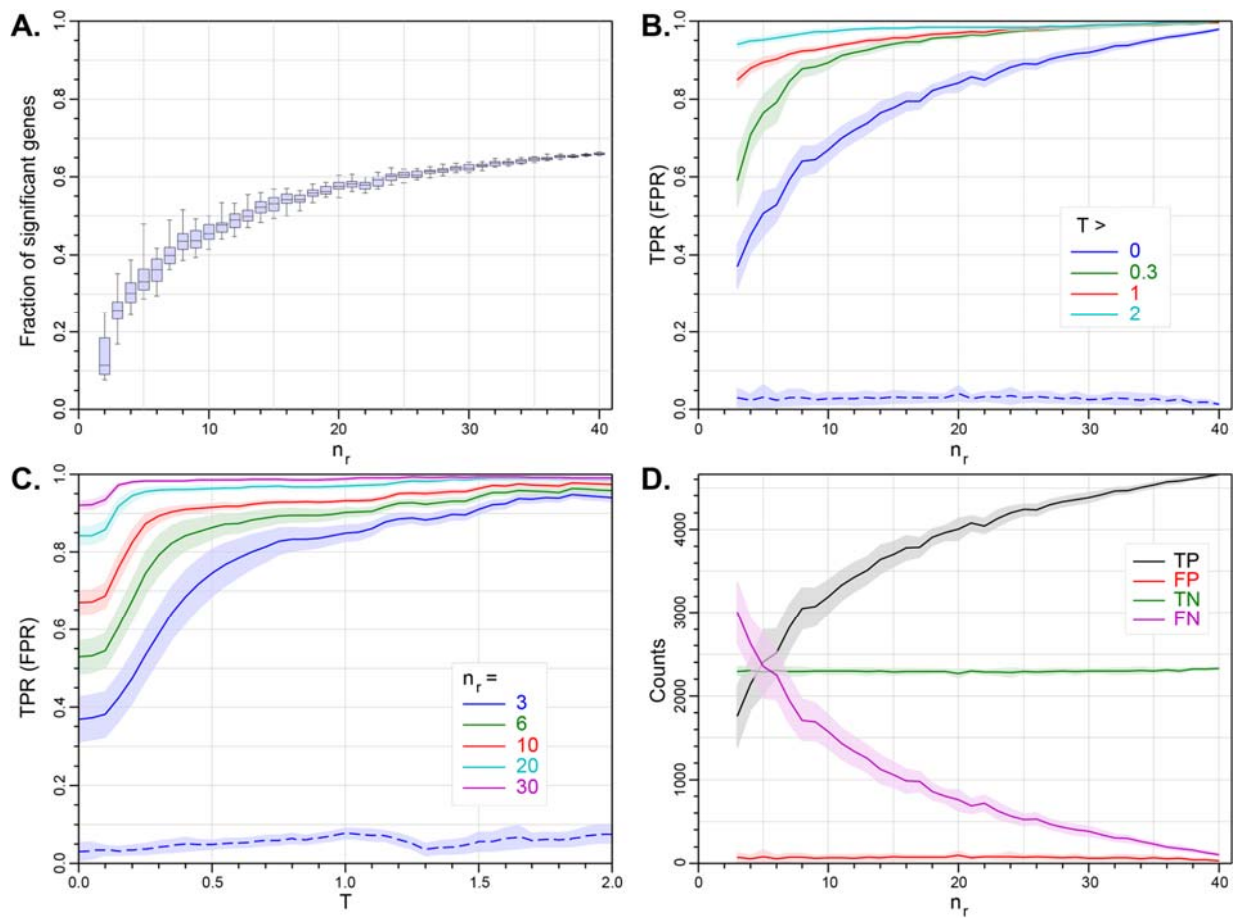

Fig S3: Statistical properties of *cuffdiff* as a function of $|\log_2 FC|$ threshold, T and the number of replicates, $n_r$. See Figure S2 for detail on each panel.

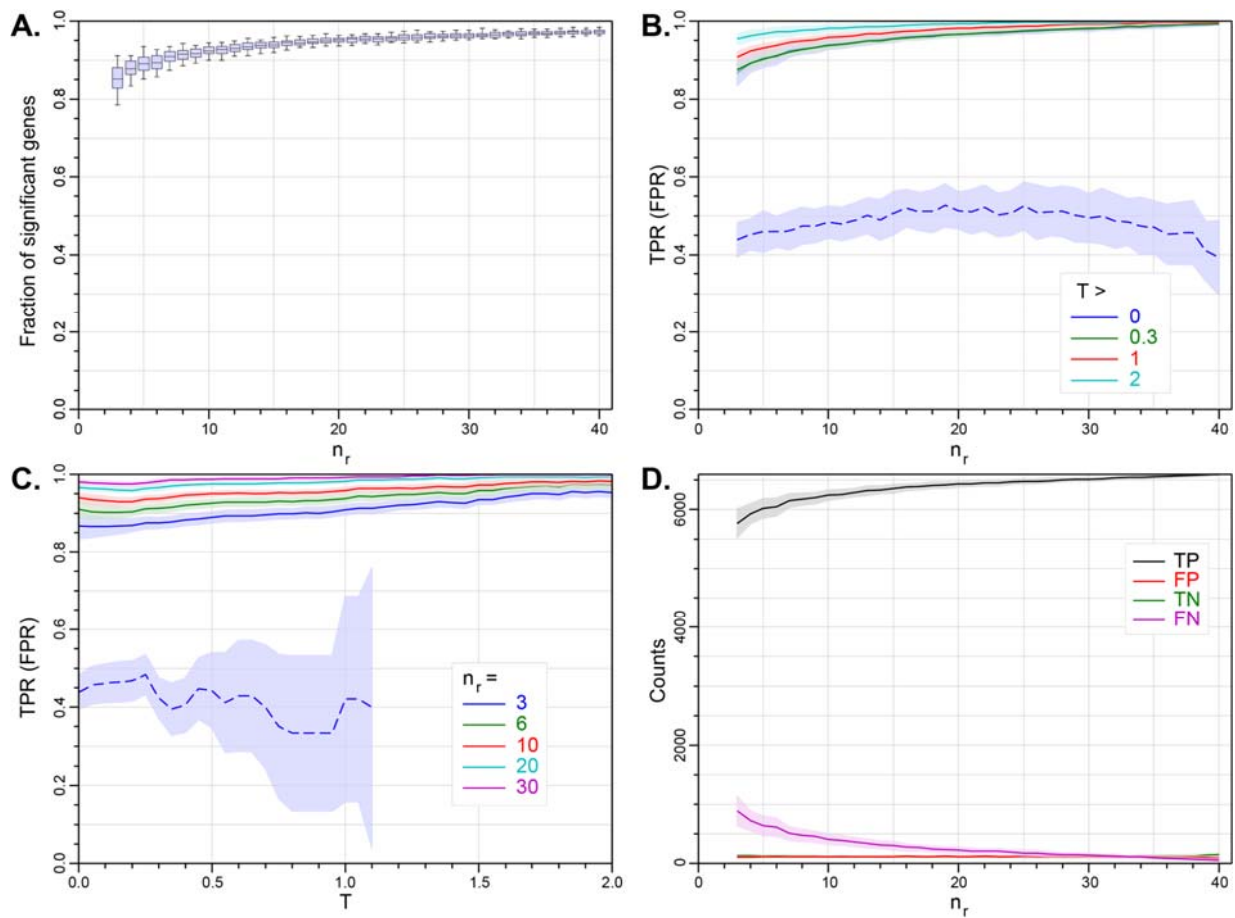

Fig S4: Statistical properties of *DEGseq* as a function of $|\log_2 FC|$ threshold, T and the number of replicates, $n_r$. See Figure S2 for detail on each panel.

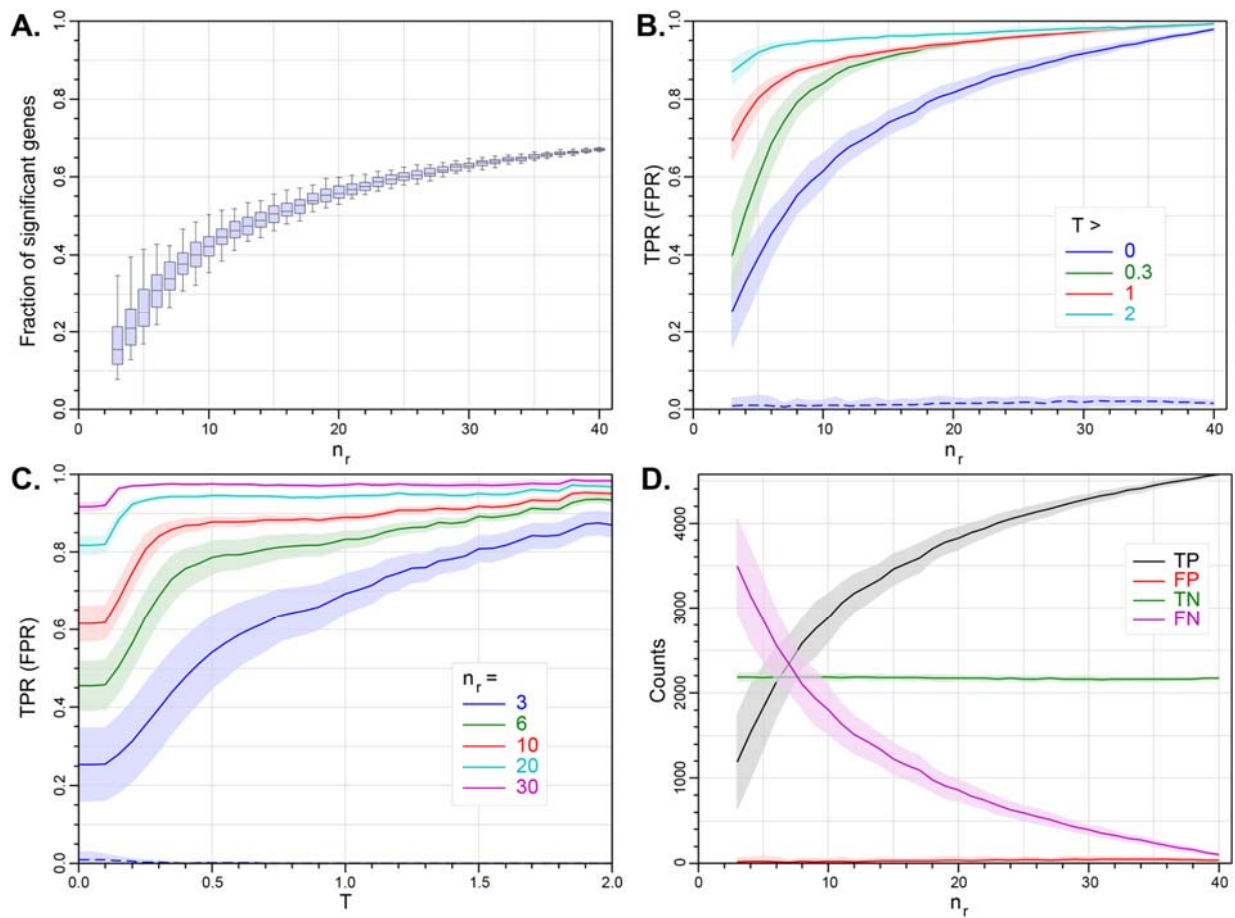

Fig S5: Statistical properties of *DEseq* as a function of $|\log_2 FC|$ threshold, T and the number of replicates, $n_r$. See Figure S2 for detail on each panel.

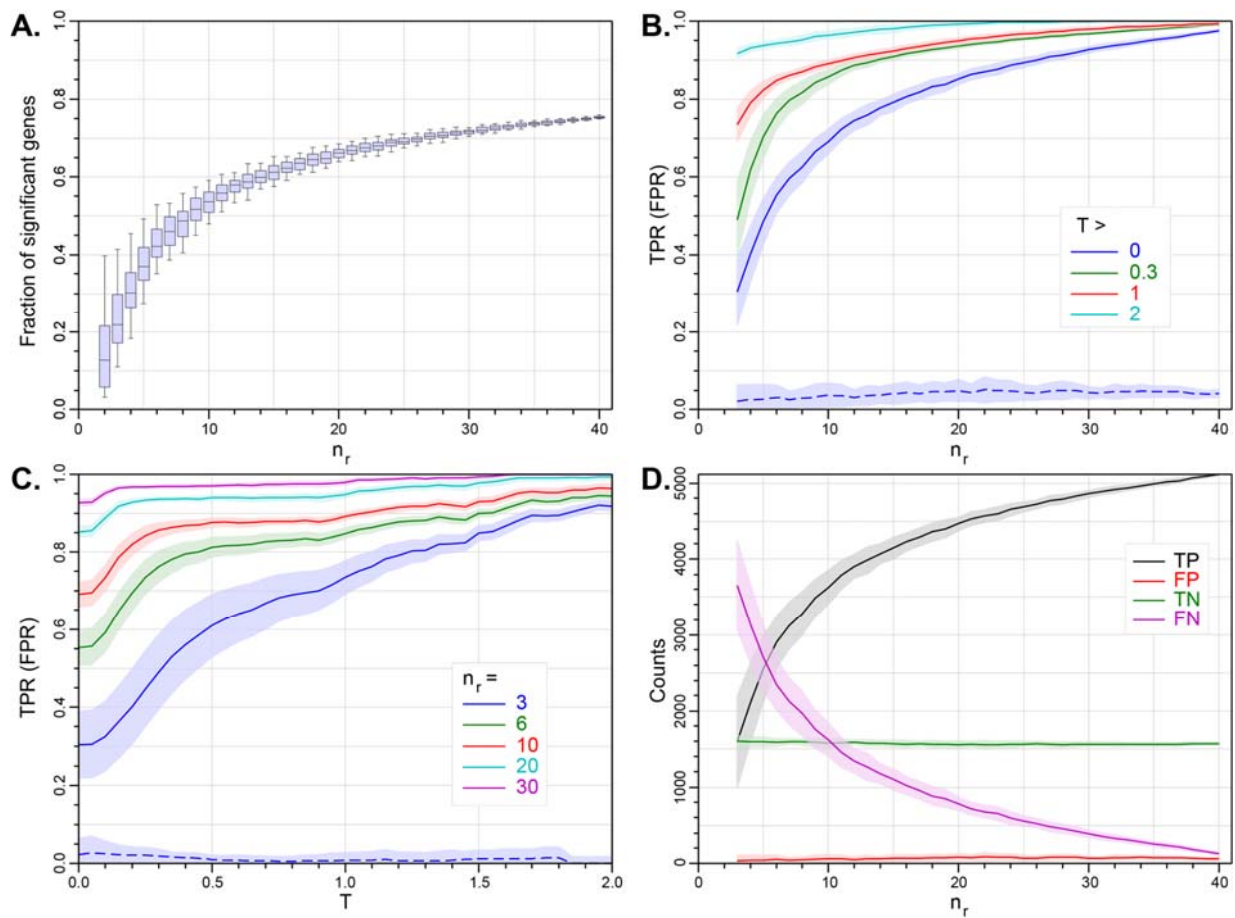

Fig S6: Statistical properties of *limma* as a function of $|\log_2 FC|$ threshold, T and the number of replicates, $n_r$. See Figure S2 for detail on each panel.

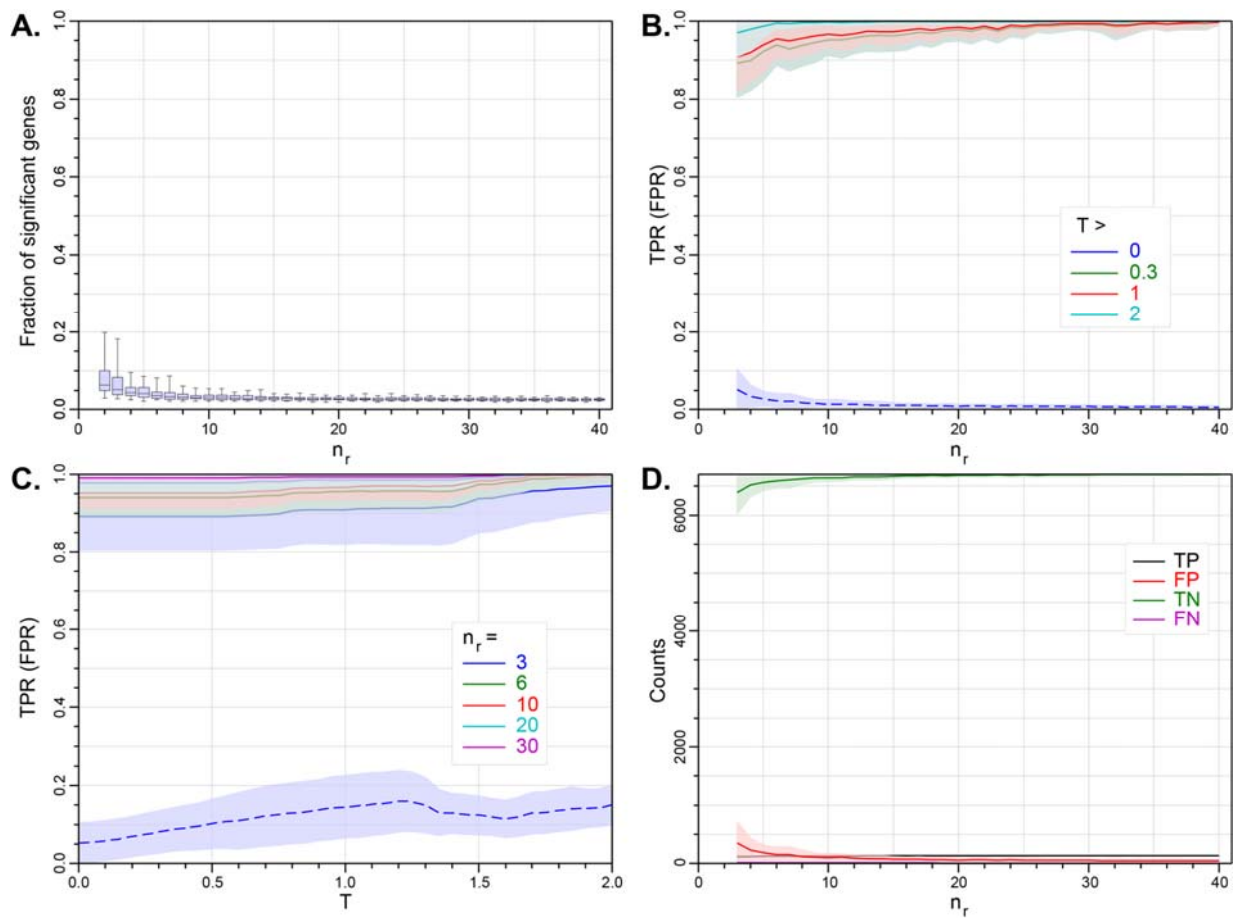

Fig S7: Statistical properties of *NOISeq* as a function of $|\log_2 FC|$ threshold, T and the number of replicates, $n_r$. See Figure S2 for detail on each panel.

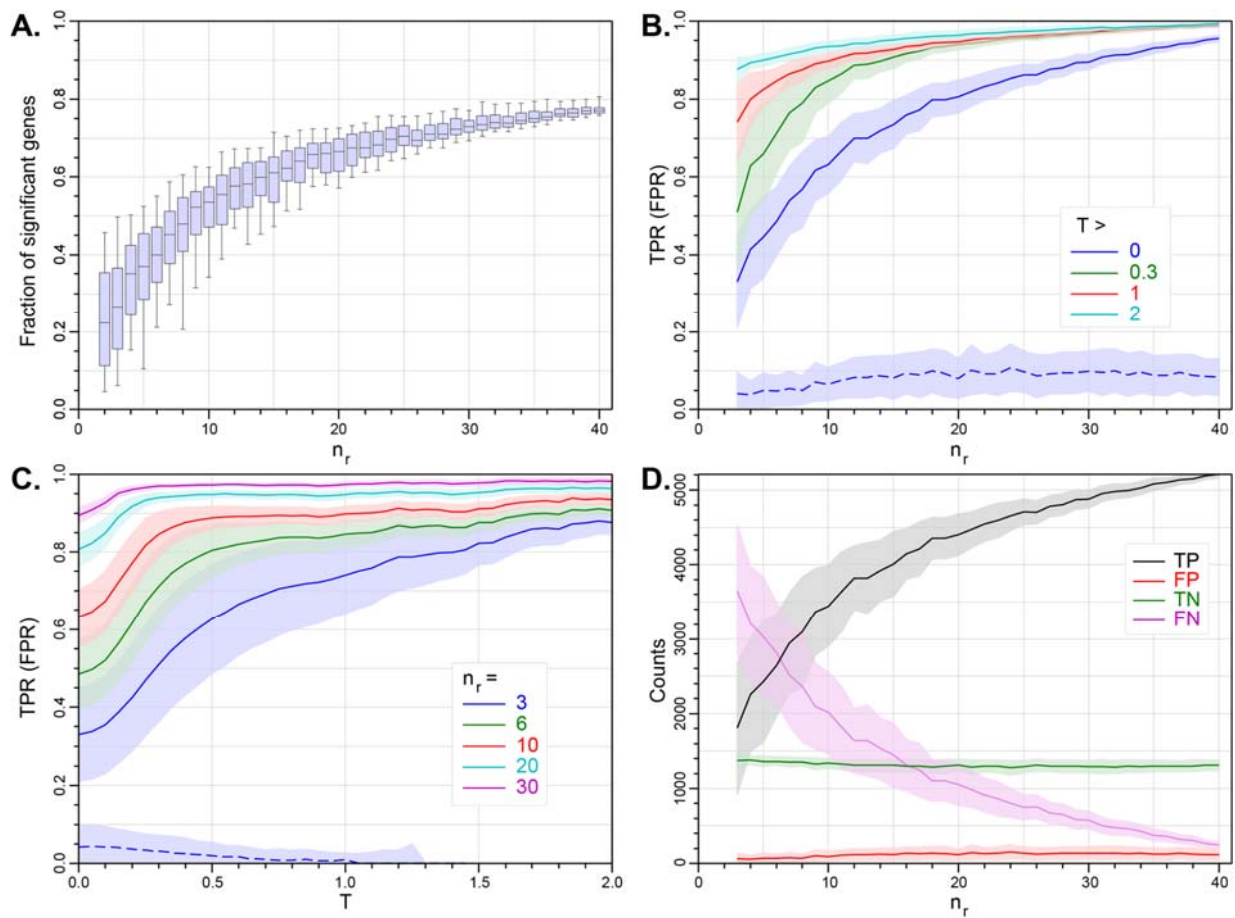

Fig S8: Statistical properties of *PoissonSeq* as a function of $|\log_2 FC|$ threshold, T and the number of replicates, $n_r$. See Figure S2 for detail on each panel.

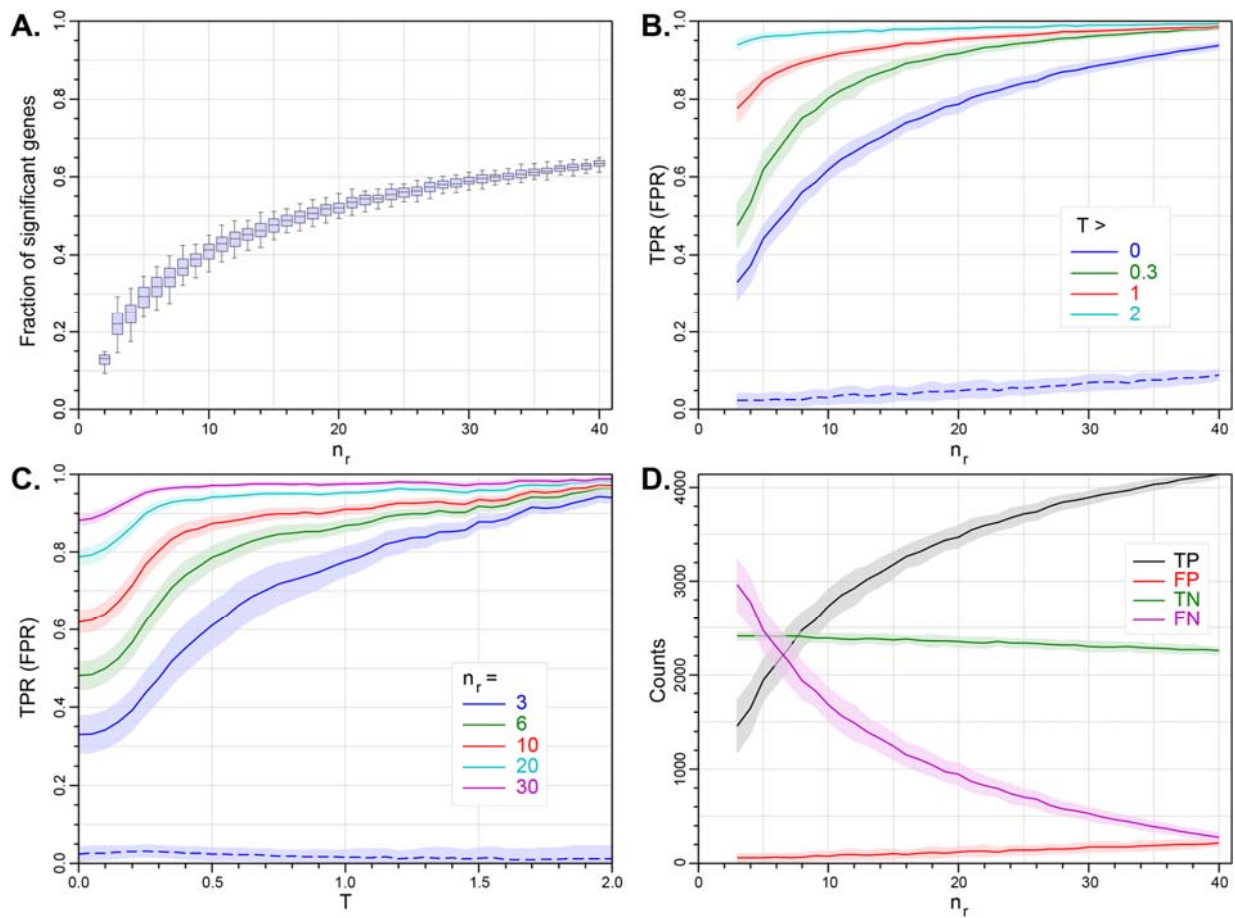

Fig S9: Statistical properties of *SAMseq* as a function of $|\log_2 FC|$ threshold, T and the number of replicates, $n_r$. See Figure S2 for detail on each panel.

31